\documentclass[aps,pra,12pt,groupedaddress,amsmath,amssymb]{revtex4}

\usepackage{bm}

\begin{document}


\mbox{}\\
\title{Affine maps of density matrices}


\author{Thomas F. Jordan}
\email{tjordan@d.umn.edu}

\affiliation{Physics Department, University of Minnesota, Duluth,
Minnesota 55812}

\date{\today}

\begin{abstract}
For quantum systems described by finite matrices, linear and affine
maps of matrices are shown to provide equivalent descriptions of
evolution of density matrices for a subsystem caused by unitary
Hamiltonian evolution in a larger system; an affine map can be
replaced by a linear map, and a linear map can be replaced by an
affine map. There may be significant advantage in using an affine
map. The linear map is generally not completely positive, but the
linear part of an equivalent affine map can be chosen to be
completely positive and related in the simplest possible way to the
unitary Hamiltonian evolution in the larger system.
\end{abstract}


\maketitle


~~~ We are accustomed to the use of linear maps of matrices to
describe evolution of density matrices in the dynamics of open
quantum systems \cite{ref1,ref2,ref3,ref4,ref5,ref6,ref7,ref8,ref9}.
It was pointed out recently \cite{ref10} that affine maps might be
used as well. I will show here that the linear and affine methods of
description are equivalent for systems described by  finite
matrices: an affine map can be replaced by a linear map, and a
linear map can be replaced by an affine map. There may be
significant advantage in using an affine map. A linear map that
describes evolution of density matrices for a subsystem caused by
unitary Hamiltonian evolution in a larger system is generally not
completely positive \cite{ref11}, but the linear part of an
equivalent affine map can be chosen to be completely positive and
related in the simplest possible way to the unitary Hamiltonian
evolution in the larger system. The equivalence demonstrated here is
for a finite-dimensional Hilbert space. It may be that one kind of
map will work where the other does not when the Hilbert space is
infinite-dimensional. There too, it may be advantageous to have two
alternatives.

~~~Consider a quantum system described by $N{\times}N$ matrices. Let
$L$ be a linear map of $N{\times}N$ matrices to $N{\times}N$
matrices; it takes each $N{\times}N$ matrix $Q$ to an $N{\times}N$
matrix  $L(Q)$. Let $K$ be an $N{\times}N$ matrix.  The map $M$ that
takes each $N{\times}N$ matrix $Q$ to the matrix
\begin{equation}
M(Q) = L(Q) + K
\end{equation}
is called $\textit{affine}$. Let $\rho$ and $\sigma$ be $N{\times}N$
density matrices and let
\begin{equation}
\tau = q\rho + (1-q)\sigma
\end{equation}
with $0<q<1$. Then $\tau$ is a density matrix and
\begin{eqnarray}
  M(\tau)& = &qL(\rho) + (1-q)L(\sigma) + K\nonumber\\
         & = &q[L(\rho) + K] + (1-q)[L(\sigma) + K]\nonumber\\
          & = &qM(\rho) + (1-q)M(\sigma).
\end{eqnarray}
If $M(\rho)$ and $M(\sigma)$ are density matrices, then $M(\tau)$ is
a density matrix and it is related to $M(\rho)$ and $M(\sigma)$ the
same as it would be if $M$ were linear. The property of linear maps
that has physical meaning for density matrices is provided by affine
maps as well.

~~~In the $N^2$-dimensional linear space of $N{\times}N$ matrices we
can find a basis of $N^2$ Hermitian matrices $F_\mu$ for $\mu$ = 0,
1, ... $N^2$-1 such that $F_0$ is 1 and

\begin{equation}
 Tr [F_\mu F_\nu] = N\delta_{\mu\nu}.
\end{equation}
If $\rho$ is a density matrix, then
\begin{equation}
 \rho =\frac{1}{N}\bigglb(1 + \sum_{\alpha=1}^{N^2-1} \langle F_\alpha \rangle F_\alpha\biggrb)
\end{equation}
with $\langle F_\alpha \rangle$ = $Tr[F_\alpha \rho]$, and

\begin{eqnarray}
 M(\rho)& =& {\frac{1}{N}\bigglb(L(1) + \sum_{\alpha=1}^{N^2-1}  \langle F_\alpha
\rangle L (F_\alpha) \biggrb) + K} \nonumber\\
 & =& {\frac{1}{N}\bigglb(L(1) + NK + \sum_{\alpha=1}^{N^2-1}  \langle F_\alpha \rangle L(
 F_\alpha)\biggrb)}.
 \end{eqnarray}
Compare this with the result of a linear map that takes each
$N{\times}N$ matrix $Q$ to an $N{\times}N$ matrix $Q'$. It gives
\begin{equation}
 \rho' =\frac{1}{N}\bigglb(1' + \sum_{\alpha=1}^{N^2-1}  \langle F_\alpha \rangle
 F_{\alpha}'\biggrb).
\end{equation}
The affine map and the linear map give the same result for all
density matrices if
\begin{equation}
1' = L(1) + NK, ~~~~~~~~~      F_{\alpha}'= L(F_{\alpha}).
\end{equation}
Specification of  a linear map means independent specification of
its action on each basis matrix 1 and $F_{\alpha}$ for $\alpha$ = 1,
2, .... $N^{2}$-1. We can choose $1'$ and $F_\alpha'$ to match any
 affine map. Conversely, we can choose $L(1)$, $K$, and
$L(F_\alpha)$  to match any linear map, but the choice of $L(1)$ and
$K$ is not unique.

~~~Specifically, suppose the density matrices are for a subsystem of
a larger system. Evolution of these density matrices caused by
unitary Hamiltonian evolution in the larger system can almost always
be described by a linear map of matrices \cite{ref11}. This linear
map is generally not completely positive, but the linear part $L$ of
the equivalent affine map is completely positive when $L(1)$ and $K$
are chosen so that $L(1)$ is 1. We can see this from the paper of
Jordan, Shaji and Sudarshan \cite{ref11}: if $L(1)$ is 1, then $L$
is their linear map with zeros for the parameters that involve mean
values of quantities for the larger system; and when these
parameters are all zero, their linear map is completely positive.
For the two-qubit example that they work out in detail, if $L(1)$ is
1 then $L$ is the linear map with $a_{1}$ and $a_{2}$ both zero;
they observe explicitly that if $a_{1}$ and $a_{2}$ are zero, the
map is completely positive for all $t$. In general, if $L(1)$ is 1
then $L$ is the linear map with the parameters $d_{\mu}$ all zero.
For any initial state of the subsystem, there is a density matrix
for an initial state of the larger system that gives zeros for the
parameters $d_{\mu}$ and is a product of the density matrix for the
subsystem and a density matrix for the other part of the larger
system; all that is required is zeros for mean values of quantities
for the other part of the larger system. When the parameters
$d_{\mu}$ are zero, the map can be obtained from unitary Hamiltonian
evolution starting from an initial product state for the larger
system. This implies that the map is completely positive. In fact,
if $L(1)$ is 1, then for each matrix $\rho$ for the subsystem,
density matrix or not,
\begin{equation}
L(\rho) = \frac{1}{M}Tr_{R}[e^{-iHt}\rho \otimes 1_{R} e^{iHt}]
\end{equation}
where $e^{-iHt}$ .. $e^{iHt}$ gives the unitary Hamiltonian
evolution of density matrices in the larger system, $R$ denotes the
other part of the larger system (the remainder or rest of the larger
system, which could be a reservoir), so $1_{R}$ is the unit matrix
for $R$ and $Tr_R$ is the trace over the states of $R$, and $M$ is
the dimension of  the Hilbert space for $R$; this holds for all the
basis matrices except 1 regardless of how $L(1)$ and $K$ are chosen,
and it holds for 1 when $L(1)$ is 1. Thus $L$ is related in the
simplest possible way to the unitary Hamiltonian evolution in the
larger system.

\end{document}